\documentclass{article}

\usepackage{amssymb,amsfonts,amsmath}
\usepackage{cite,enumerate,float}
\usepackage{color}
\usepackage{tikz}
\usetikzlibrary{arrows,snakes,backgrounds}
\usepackage{dsfont}

\usepackage{ytableau}
\ytableausetup{boxsize=0.75em}

\usepackage[pdftex]{pict2e}

\def\be{\begin{eqnarray}}
\def\ee{\end{eqnarray}}
\def\nn{\nonumber}

\def\Tr{{\rm Tr}\,}

\def\l[{\phantom.[}

\def\tZ{\tilde Z}
\def\tF{\tilde F}

\def\BS{\bar{\cal S}}

\def\picBdd{
\put(0,0){\line(1,0){100}}
\put(0,0){\line(0,1){30}}
\put(0,30){\line(1,0){100}}
\put(100,0){\line(0,1){30}}
}

\def\picBzz{
\put(0,0){\line(1,0){100}}
\put(0,0){\line(0,1){20}}
\put(0,20){\line(1,0){100}}
\put(100,0){\line(0,1){20}}
}

\def\picdz{
\put(0,0){\line(1,0){20}}
\put(0,0){\line(0,1){30}}
\put(20,0){\line(0,1){20}}
\put(0,30){\line(1,0){10}}
\put(10,20){\line(1,0){10}}
\put(10,20){\line(0,1){10}}
}

\def\picdd{
\put(0,0){\line(1,0){20}}
\put(0,0){\line(0,1){30}}
\put(0,30){\line(1,0){20}}
\put(20,0){\line(0,1){30}}
}

\def\piczz{
\put(0,0){\line(1,0){20}}
\put(0,0){\line(0,1){20}}
\put(0,20){\line(1,0){20}}
\put(20,0){\line(0,1){20}}
}

\def\picde{
\put(0,0){\line(1,0){20}}
\put(0,0){\line(0,1){30}}
\put(20,0){\line(0,1){10}}
\put(0,30){\line(1,0){10}}
\put(10,10){\line(1,0){10}}
\put(10,10){\line(0,1){20}}
}

\def\picd{
\put(0,0){\line(1,0){10}}
\put(0,0){\line(0,1){30}}
\put(0,30){\line(1,0){10}}
\put(10,0){\line(0,1){30}}
}

\def\picze{
\put(0,0){\line(1,0){20}}
\put(0,0){\line(0,1){20}}
\put(20,0){\line(0,1){10}}
\put(0,20){\line(1,0){10}}
\put(10,10){\line(1,0){10}}
\put(10,10){\line(0,1){10}}
}

\def\picz{
\put(0,0){\line(1,0){10}}
\put(0,0){\line(0,1){20}}
\put(0,20){\line(1,0){10}}
\put(10,0){\line(0,1){20}}
}

\def\picee{
\put(0,0){\line(1,0){20}}
\put(0,0){\line(0,1){10}}
\put(0,10){\line(1,0){20}}
\put(20,0){\line(0,1){10}}
}

\def\pice{
\put(0,0){\line(1,0){10}}
\put(0,0){\line(0,1){10}}
\put(0,10){\line(1,0){10}}
\put(10,0){\line(0,1){10}}
}

\hoffset= - 1.0in         

\begin{document}

\hfill ITEP/TH-29/16

\hfill IITP/TH-20/16

\bigskip

\centerline{\Large{
Factorization of differential expansion for non-rectangular representations
}}

\bigskip

\centerline{\bf  A.Morozov }

\bigskip

{\footnotesize
\centerline{{\it
ITEP, Moscow 117218, Russia}}

\centerline{{\it
Institute for Information Transmission Problems, Moscow 127994, Russia
}}

\centerline{{\it
National Research Nuclear University MEPhI, Moscow 115409, Russia
}}
}

\bigskip

\centerline{ABSTRACT}

\bigskip

\noindent
{\footnotesize
Factorization of the differential expansion (DE) coefficients
for colored HOMFLY-PT polynomials of antiparallel double braids,
discovered in \cite{rectwist}
in the case of rectangular representations $R$, is extended to
the first non-rectangular representations $R=[2,1]$ and $R=[3,1]$.
This increases chances that such factorization will take place for generic $R$,
thus fixing the shape of the DE.
We illustrate the power of the method by conjecturing the DE-induced expression
for double-braid polynomials for all $R=[r,1]$.
In variance with rectangular case, the knowledge for double braids
is not fully sufficient to deduce the exclusive Racah matrix $\bar S$ --
the entries in the sectors with non-trivial multiplicities sum up
and remain unseparated.
Still a considerable piece of the matrix is extracted directly
and its other elements can be found by solving the unitarity constraints.
}

\bigskip

\section{Introduction}

Wilson loop averages in $3d$ Chern-Simons theory \cite{Wit}
\be
{\cal H}_R^{\cal K}
= \ \left< \Tr\!_R \ P\exp\left( \oint_{\cal K} {\cal A}\right)\right>
\label{HR}
\ee
are exactly calculable and provide an important
set of examples for non-perturbative quantum field theory (QFT).
At this moment the calculations can be most effectively performed
with the help of the modified Reshetikhin-Turaev method \cite{RT}-\cite{mmms21},
where the answers are certain combinations of various Racah matrices
($6j, 9j, \ldots$ -symbols).
The problem is, however, that these matrices are not known for
most representations $R$, and their direct evaluation is far beyond
modern computer capacities.
Also, most valuable is not the answer for particular representation $R$,
but its analytic dependence on the representation -- and here even the
absolute success of calculation for particular $R$ (which we are very
far from) would not be sufficient.
Still, there is a considerable progress in the field during the last decade,
and it reveals the interesting properties of the quantities ${\cal H}_R^{\cal K}$,
which in no way follow from their definition (\ref{HR}) --
and imply the existence of some complementarity (dual) descriptions,
which still remain to be found.

\bigskip

The puzzling properties of ${\cal H}_R^{\cal K}$ include:

\bigskip

$\bullet$ ${\cal H}_R^{\cal K}$ are polynomials in two non-perturbative variables,
$q = \exp\left(\frac{2\pi i}{g^2+N}\right)$ and $A=q^N$, which are made from the
coupling constant and parameter of the gauge group $SL(N)$.
This is a long known fact, coming from identification \cite{Wit} of
${\cal H}_R^{\cal K}$ with the HOMFLY-PT polynomials \cite{knotpols},
but its exact meaning in QFT remains obscure.

$\bullet$ Coefficients of the polynomials are integer, what is explained in
alternative Khovanov-Rozansky approach \cite{KhR}, however, its QFT interpretation
is still unavailable.
Moreover, this approach is not yet developed for non-trivial representations $R$,
and it does not  explain the more delicate integrality properties
\cite{OV,integrality}
of Ooguri-Vafa sums $\sum_R {\cal H}_R\chi_R$ and their loop expansions.

$\bullet$ Knots can be glued from simpler components, and this provides a new
description of knot polynomials -- in terms of some effective gauge
invariant field theory. This line of reasoning is so far developed
\cite{mmmrsv1}-\cite{mmms21}
for a family of arborescent knots \cite{arbotexts},
which are distinguished because in this case the problem is reduced to just
two types of "exclusive" Racah matrices $S$ and $\bar S$
(which, however, depend on representation $R$ and are still very non-trivial
to calculate).

$\bullet$ Vogel's universality \cite{Vogel} works perfectly well for knot polynomials
\cite{univMMM,univMM}:
while particular dimensions in the $E_8$-sector of representation theory
are transcendental in the $u,v,w$-variables, they combine into Vieta-like
rational combinations in the expressions for ${\cal H}_R^{\cal K}$
for $R$ which are descendants of the adjoint representation.
Perhaps, this is not too surprising, because universality was actually inspired
by knot-theory considerations -- and it is now getting clear that knot theory
distinguishes a "healthy" part of representation theory,   which includes
the exclusive Racah matrices $S$ and $\bar S$ (they are currently known
in the universal form for the adjoint representation itself \cite{univMM}).

$\bullet$ As functions of $R$, HOMFLY-PT polynomials satisfy non-trivial
difference equations \cite{Gar,knotpolseqs}
-- so far fully described only for symmetric representations $R=[r]$ and for
particular knots, where the full $r$-dependence is known from \cite{IMMMfe,sympols}.

$\bullet$ The HOMFLY-PT polynomials possess a non-trivial structure of
differential expansion (DE) \cite{IMMMfe}, \cite{evo}-\cite{Konodef} --
which probably reflects the duality between
Reshetikhin-Turaev and Khovanov-Rozansky approaches, though both are not
yet formulated in such a way that this statement can be made explicit.

\bigskip

At this moment DE is deepest structure, found in knot polynomials,
and the present paper is devoted to a new progress in its investigation.

\bigskip

DE controls the dependence of knot polynomials on representation $R$
and can be used to characterize (and further -- classify) the
{\it complexity} of knots -- already the DE defect of \cite{Konodef}
(occasionally equal to minus one plus degree of the fundamental
Alexander polynomial)
seems to be a much better characteristic than the usual minimal-crossing number.
DE is quite a powerful tool -- originally it was used in
\cite{IMMMfe,sympols} to find the "exclusive" Racah matrices $S$ and $\bar S$
for all  symmetric and antisymmetric representations $R=[r], [1^r]$,
and recently the method was extended in \cite{rectwist} to generic rectangular $R$
(though the calculation is not yet completed in generic case).
This power comes from promoting the observation about available
hardly-calculated examples for particular $R$ to all $R$
of a certain type --
and thus obtaining (conjecturing) the statements far beyond the domain
of direct calculability.
The main problem with DE at {\it this} moment, after its tremendous success
with rectangular $R$, is the lasting difficulty for non-rectangular $R$,
beginning already at the simplest level of $R=[2,1]$,
see \cite{de21,twevo21} and \cite{mmms21} for a number of previous attempts.
With the new knowledge and insight from \cite{rec41,konotwist}
we now manage to resolve this $[2,1]$-problem --
and this is what the present paper is about.

Namely, we suggest, that DE for $R=[2,1]$ involves {\it two} new
structures, as compared to the case of symmetric $R$, which we denote
$F_{[2,1]}$ and $\tF_{[2,1]}$, and non of them is reduced to the
previously known $F_{[3]}$ -- contrary to what was assumed so far.
Spectacularly, this conjecture appears consistent with another conjecture --
about factorization of DE for double braids \cite{rectwist},
and this looks absolutely non-trivial and extremely restrictive,
leaving practically no doubts in validity of the both, at least for $R=[2,1]$.
We also provide a simple example of the defect-zero knot $9_{46}$
which is beyond the double-braid family.

Once the case of $R=[2,1]$ is understood,
the way is open to other non-rectangular representations,
though it is neither fully straightforward nor easy.
Still as an illustration that now this is  doable,
we provide DE for the next representation $R=[3,1]$.

\section{Differential expansion and its factorization for double braids}

In this letter we assume the familiarity with the summary \cite{konotwist}
of the recent developments about DE for rectangular $R$
and provide only the new details,
needed beyond rectangular representations.
The short list of abbreviations, used in the theory of knot polynomials
include  $\{x\} = x-x^{-1}$, the quantum numbers $[n]=\frac{\{q^n\}}{\{q\}}$
and the "differentials" $D_n=\{Aq^n\}$.

{\bf Differential expansion}, as we currently understand it, decomposes
colored HOMFLY-PT polynomial and separate representation (color)
and braid dependencies in the following way:
\be
\boxed{
{\cal H}_R^{\cal K} = \sum_{\lambda} C_R^{\lambda}(q) \cdot Z_R^{\lambda}(A,q)\cdot
F_\lambda^{\cal K}(A,q)
}
\label{de}
\ee
Combinatorial coefficients $C(q)$ can contain denominators, made from $q$-numbers.
The Laurent-polynomial $Z$-factors depend on the defect of the knot ${\cal K}$, or,
more precisely,   $F^{\cal K}$ contain knot-independent $D$-factors,
which can be absorbed into $Z$.
In this paper we consider the knots with defect zero.
All the $F$-factors also are Laurent polynomials,
moreover,  for the three distinguished knots
\be
F_\lambda^{unknot}=0,\ \ \ \ \
 F_\lambda^{4_1}=1,\ \ \ \ \
 F_\lambda^{3_1}=(-)^{\alpha} q^{\beta} A^{\gamma}
\label{Fdist}
 \ee
 with some $\lambda$-dependent integers $\alpha,\beta,\gamma$.

For {\bf rectangular} $R=[r^s]$ the sum goes over $\lambda$,
which are sub-diagrams of $R$.
For non-rectangular $R$ there are additional ("anomalous")
 contributions, which can {\it not} be associated {\it just} with the sub-diagrams
 $\lambda\subset R$ and which depend also on the eigenvalues,
 {\it not} directly associated with sub-diagrams $\mu \subset \lambda$.
 These additional contributions, which we denote by tilde,
 however,  come with additional factors of $\{q\}^4$.
 Our notation is not too informative, what reflects the lack of a true
 understanding of the phenomenon -- the goal of the present paper is mostly
 to put it to light and describe, not yet to fully explain.

{\bf Double braid} of type, relevant for our purposes in this paper
is shown in the picture:

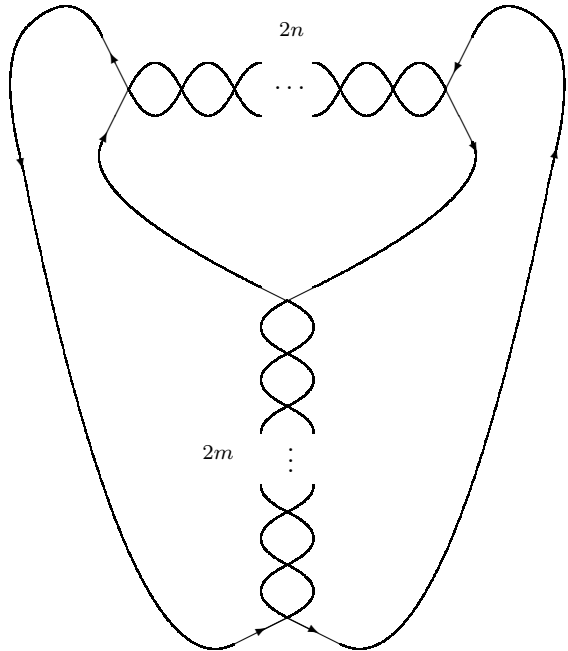
\begin{picture}(200,280)(-230,-230)
\qbezier(-40,0)(-50,20)(-60,0)
\qbezier(-40,0)(-50,-20)(-60,0)
\qbezier(-20,0)(-30,20)(-40,0)
\qbezier(-20,0)(-30,-20)(-40,0)
\qbezier(-20,0)(-15,10)(-10,10)
\qbezier(-20,0)(-15,-10)(-10,-10)
\put(-5,0){\mbox{$\ldots$}}
\qbezier(10,10)(15,10)(20,0)
\qbezier(10,-10)(15,-10)(20,0)
\qbezier(20,0)(30,20)(40,0)
\qbezier(20,0)(30,-20)(40,0)
\qbezier(40,0)(50,20)(60,0)
\qbezier(40,0)(50,-20)(60,0)
\put(-60,0){\line(-1,2){10}}
\put(-60,0){\line(-1,-2){10}}
\put(60,0){\line(1,2){10}}
\put(60,0){\line(1,-2){10}}
\qbezier(0,-80)(-20,-90)(0,-100)
\qbezier(0,-80)(20,-90)(0,-100)
\qbezier(0,-100)(-20,-110)(0,-120)
\qbezier(0,-100)(20,-110)(0,-120)
\qbezier(0,-120)(-10,-125)(-10,-130)
\qbezier(0,-120)(10,-125)(10,-130)
\put(0,-145){\mbox{$\vdots$}}
\qbezier(0,-160)(-10,-155)(-10,-150)
\qbezier(0,-160)(10,-155)(10,-150)
\qbezier(0,-160)(-20,-170)(0,-180)
\qbezier(0,-160)(20,-170)(0,-180)
\qbezier(0,-180)(-20,-190)(0,-200)
\qbezier(0,-180)(20,-190)(0,-200)
\put(0,-80){\line(-2,1){10}}
\put(0,-80){\line(2,1){10}}
\put(0,-200){\line(-2,-1){10}}
\put(0,-200){\line(2,-1){10}}
\put(0,-200){\line(-2,-1){20}}
\put(0,-200){\line(2,-1){20}}
\qbezier(-10,-75)(-80,-40)(-70,-20)
\qbezier(10,-75)(80,-40)(70,-20)
\put(-10,-205){\vector(2,1){2}}
\put(10,-205){\vector(2,-1){2}}
\put(-65,10){\vector(-1,2){2}}
\put(65,10){\vector(-1,-2){2}}
\put(-70,-20){\vector(1,2){2}}
\put(70,-20){\vector(1,-2){2}}
\put(-3,20){\mbox{\footnotesize$2n$}}
\put(-32,-140){\mbox{\footnotesize $2m$}}
%
\qbezier(-70,20)(-80,40)(-97,25)
\qbezier(-97,25)(-111,13))(-100,-30)
\qbezier(-100,-30)(-60,-230)(-20,-210)
\qbezier(70,20)(80,40)(97,25)
\qbezier(97,25)(111,13))(100,-30)
\qbezier(100,-30)(60,-230)(20,-210)
\put(-102,-22){\vector(1,-4){2}}
\put(100,-30){\vector(1,4){2}}
\end{picture}

\noindent
{\bf Twist knots} form a particular subset of this two-parametric family
with $n=2$ and arbitrary $m$.
For rectangular $R$  in this case every $F_\lambda^{(m)}$
is a linear combination of powers of the eigenvalues $\Lambda_\mu^{2m}$,
which are labeled by sub-diagrams $\mu\subset \lambda$.
As an amusing side remark, all $F_\lambda$ seem to be polynomials
with {\it positive} integer coefficients --
this emphasizes their relation to superpolynomials, first suggested in the
original \cite{IMMMfe}.

For double braids $F_\lambda^{(m,n)}$ are bilinear combinations of
$\Lambda_\mu^{2m}\Lambda_{\mu'}^{2n}$.
However,
{\bf factorization} of HOMFLY-PT polynomials for {\bf double braids},
discovered in \cite{rectwist}, reduces them to those for {\bf twist knots} --
but only if both are realized by their differential expansions:
factorization states that
\be
\boxed{
F_\lambda^{(m,n)} \sim F_\lambda^{(m)}F_\lambda^{(n)}
}
\ee
In this paper we extend this conjecture from rectangular
representations $R$ to the simplest non-rectangular $R=[2,1]$ and $R=[3,1]$.

\section{Representation $R=[1,1]$}

As a by-now-elementary starting point, we remind the DE formulas from \cite{IMMMfe}
and \cite{rectwist}
for the antisymmetric representation $R=[1,1]$:
\be
{\cal H}_{[1,1]}^{(m,n)} = 1
+ Z^{[1]}_{[1,1]}\cdot \frac{F_{[1]}^{(m)}F_{[1]}^{(n)}}{F_{[1]}^{(1)}F_{[1]}^{(-1)}}
 +   Z^{[1,1]}_{[1,1]}\cdot
\frac{F_{[1,1]}^{(m)}F_{[1,1]}^{(n)}}{F_{[1,1]}^{(1)}F_{[1,1]}^{(-1)}}
\label{H11}
\ee
with the $Z$-factors
\be
Z^{[1]}_{[1,1]}=D_1D_{-3} + D_1D_{-1} = [2]\,D_1D_{-2}, \ \ \ \ \ \ \
Z^{[1,1]}_{[1,1]}= D_1\underline{D_0}D_{-2}D_{-3}
\ee
Underlined is the differential, which is omitted from the differential expansion
for knots with  defects greater than zero, see \cite{Konodef}.
The two $F$-functions can be found in the list (\ref{Ftwist21}) below.

\section{Representation $R=[2,1]$}

\bigskip

Contributing to the DE (\ref{de}) in this case are seven different
Young diagrams from $R\otimes R = [2,1]\otimes \overline{[2,1]}$,
of which five are naturally labeled by sub-diagrams
$\lambda$ of $R=[2,1]$ itself:

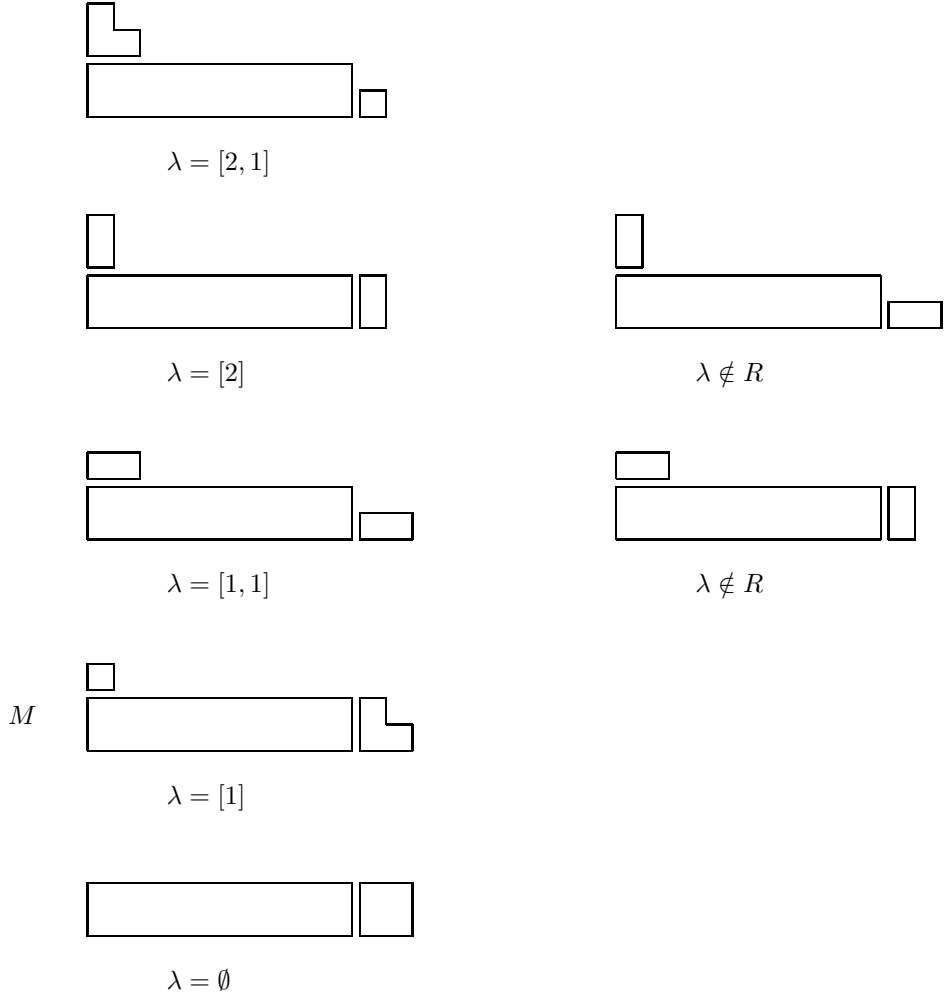
\begin{picture}(500,395)(-50,-340)
\picBzz
\put(0,23){\picze}
\put(103,0){\pice}
\put(30,-20){\put(0,0){\mbox{$\lambda=[2,1]$}}}
\put(0,-80){
\picBzz
\put(0,23){\picz}
\put(103,0){\picz}
\put(30,-20){\put(0,0){\mbox{$\lambda=[2]$}}}
}

\put(200,-80){
\picBzz
\put(0,23){\picz}
\put(103,0){\picee}
\put(30,-20){\put(0,0){\mbox{$\lambda\notin R$}}}
}
\put(0,-160){
\picBzz
\put(0,23){\picee}
\put(103,0){\picee}
\put(30,-20){\put(0,0){\mbox{$\lambda=[1,1]$}}}
}
\put(200,-160){
\picBzz
\put(0,23){\picee}
\put(103,0){\picz}
\put(30,-20){\put(0,0){\mbox{$\lambda\notin R$}}}
}
\put(0,-240){
\picBzz
\put(0,23){\pice}
\put(103,0){\picze}
\put(30,-20){\put(0,0){\mbox{$\lambda=[1]$}}}
\put(-30,10){\mbox{$M$}}
}
\put(0,-310){
 \picBzz
\put(0,23){ }
\put(103,0){\piczz}
\put(30,-20){\put(0,0){\mbox{$\lambda=\emptyset$}}}
}
\end{picture}
These five "regular" diagrams are shown in the left column.
Somewhat miraculously, the two "anomalous" diagrams in the right column
have identical dimensions and Casimir eigenvalues --
thus it is not a big surprise that they provide a single
(rather than two separate) contributions to the differential expansion.
Note that double-braid {\it factorization} is bilinear and
thus very sensitive to the difference
between single and two separate contributions.
As to the {\it multiplicity}, in the case of $R=[2,1]$
it appears only for $\lambda=[1]$.
Surprisingly or not,
it does not show up in the differential expansion.

Making use of the results of \cite{GJ} for the $[2,1]$ Racah matrices
and of \cite{twevo21} for the evolution of the $[2,1]$-colored twisted knots,
by a tedious trial-and-error attempt, we discover the following differential
expansion, which nicely fits all the known results for
double-braid $[2,1]$-colored HOMFLY-PT from \cite{knotebook}:

\be
{\cal H}_{[2,1]}^{(m,n)} = 1
+ Z^{[1]}_{[2,1]}\cdot \frac{F_{[1]}^{(m)}F_{[1]}^{(n)}}{F_{[1]}^{(1)}F_{[1]}^{(-1)}}
+ \frac{[3]}{[2]}\cdot Z^{[2]}_{[2,1]}\cdot
\frac{F_{[2]}^{(m)}F_{[2]}^{(n)}}{F_{[2]}^{(2)}F_{[1]}^{(-1)}}
+ \frac{[3]}{[2]}\cdot Z^{[1,1]}_{[2,1]}\cdot
\frac{F_{[1,1]}^{(m)}F_{[1,1]}^{(n)}}{F_{[1,1]}^{(1)}F_{[1,1]}^{(-1)}}
+ Z^{[2,1]}_{[2,1]}\cdot
\frac{F_{[2,1]}^{(m)}F_{[2,1]}^{(n)}}{F_{[2,1]}^{(1)}F_{[2,1]}^{(-1)}} + \nn \\
+ {\tZ}^{[2,1]}_{[2,1]}\cdot
\frac{\tF_{[2,1]}^{(m)}\tF_{[2,1]}^{(n)}}{\tF_{[2,1]}^{(1)}\tF_{[2,1]}^{(-1)}}
\label{H21de}
\ee
with the knot-independent $Z$-factors
\be
Z^{[1]}_{[2,1]}=D_3D_{-3} + D_2D_0+D_0D_{-2}, \ \ \ \ \ \ \ \
Z^{[2]}_{[2,1]}= D_3D_2\underline{D_0}D_{-2}, \ \ \ \ \ \ \ \
Z^{[1,1]}_{[2,1]}= D_2\underline{D_0}D_{-2}D_{-3} \nn \\
Z^{[2,1]}_{[2,1]}= D_3D_2\underline{D_1D_{-1}}D_{-2}D_{-3}, \ \ \ \ \ \ \ \ \ \ \ \ \
\tZ^{[2,1]}_{[2,1]}= -[3]^2\{q\}^4 D_2D_{-2}\ \ \ \ \ \ \ \ \ \ \ \
\label{Zf21}
\ee
The first three $Z$-factors are known since \cite{IMMMfe}
from DE in symmetric representations,
the last two are new.
The knot-dependent $F$-factors for twist family are
\be
F_{[1]}^{(m)} = A\cdot\left(\frac{\Lambda_0^{2m}}{D_0}
- \frac{\Lambda_1^{2m}}{D_0}\right)\nn \\
F_{[2]}^{(m)} = q\cdot A^2\cdot\left(\frac{\Lambda_0^{2m}}{D_1D_0}
- [2]\cdot \frac{\Lambda_1^{2m}}{D_2D_0}+ \frac{\Lambda_{2}^{2m}}{D_2D_1}\right)\nn \\
F_{[1,1]}^{(m)} = q^{-1}\cdot A^2\cdot\left(\frac{\Lambda_0^{2m}}{D_0D_{-1}}
- [2]\cdot \frac{\Lambda_1^{2m}}{D_0D_{-2}}
+ \frac{\Lambda_{11}^{2m}}{D_{-1}D_{-2}}\right)\nn \\
F_{[2,1]}^{(m)} =  A^3\cdot\left(\frac{\Lambda_0^{2m}}{D_1D_0D_{-1}}
- [3]\cdot \frac{\Lambda_1^{2m}}{D_2D_0D_{-2}}
+ \frac{[3]}{[2]}\cdot \frac{\Lambda_{2}^{2m}}{D_2D_1D_{-2}}
+ \frac{[3]}{[2]}\cdot \frac{\Lambda_{11}^{2m}}{D_2D_{-1}D_{-2}}
-   \frac{\Lambda_{21}^{2m}}{D_2D_0D_{-2}} \right)\nn \\
\tF_{[2,1]}^{(m)} = A^3\cdot\left(
\frac{\Lambda_0^{2m}}{D_1D_0D_{-1}}
- \frac{[4]}{[2]}\cdot\frac{\Lambda_1^{2m}}{D_2D_0D_{-2}}
-\frac{1}{[2]^2\{q\}^2}\cdot
\left( \frac{D_3\,\Lambda_{2}^{2m}}{D_2D_1}
-  \frac{2\tilde\Lambda_2^{2m}}{D_0}
 + \frac{D_{-3}\,\Lambda_{11}^{2m}}{D_{-1}D_{-2}}\right)
\right)
\label{Ftwist21}
\ee
with $\Lambda_0^2=1,\ \  \Lambda_1^2 = A^2, \ \
\underline{\tilde\Lambda_2^2 = A^4},\ \  \Lambda_{2}^2 = q^4A^4,\ \  \Lambda_{11}^2 = q^{-4}A^4, \ \
\Lambda_{21}^2 = A^6$.
Underlined is the eigenvalue, associated with the two "anomalous" diagrams in the
second column. Wherever possible we put tildes over dimensions rather than
representation labels (subscripts) to make them better visible.
Note that $\tF_{[2,1]}^{[2,1]}$ does not depend on $\Lambda_{21}$.

\section{Comments and checks}

The formula (\ref{H21de}) differs from most previous suggestions
about the differential expansion for ${\cal H}_{[2,1]}$,
but this time it works nicely for all available twist and double-braid answers.
A few additional checks/comments are now in order.

\bigskip

{\bf 5.1.} It is easy to see, that the
{\bf Alexander polynomial} ${\rm Al}^{\cal K}(q) = {\cal H}^{\cal K}(A=1,q)$
 is strongly affected by the last term with  $\tZ_{[2,1]}^{[2,1]}$
 and $\tF_{[2,1]}^{[2,1]}$ in (\ref{H21de}),
and this provides an additional non-trivial check of our formulas -- because is can
be performed for arbitrarily large values of $m$ and $n$, where alternative answers for
the $[21]$-colored polynomials are not available.
The thing to check is that  the Alexander polynomial an arbitrary single-hook
representation $R$ satisfies \cite{mmAl}
\be
{\rm Al}_R(q) = {\rm Al}_{[1]}(q^{|R|})\ \ \ \ \ {\rm for \ single-hook \ }
R = [p,1^q]
\label{Alred}
\ee

\bigskip

{\bf 5.2.} As explained in \cite{rectwist}, the differential expansion for double braids
contains important information about the {\bf Racah matrix} $\bar S$,
which allows to fully extract it for arbitrary rectangular representation
(the practical obstacle there is incomplete knowledge of the $F_\lambda$-functions
for Young diagrams $\lambda$ with more than two columns).
However, for non-rectangular diagrams such extraction is not fully possible:
contributions from representations with multiplicities are summed and additional
effort is needed to separate them.
In the case of $R=[21]$ this looks as follows.
Just as in \cite{rectwist},
one can easily deduce all the elements of the matrix $\bar{\cal S}_{ab}$ from
\be
H_{[2,1]}^{(m,n)} = \sum_{a,b=0}^5
\frac{\sqrt{d_ad_b}}{d_{[2,1]}}\cdot \bar {\cal S}_{ab}
\cdot \Lambda_a^{2m}\Lambda_b^{2n}
\label{21calS}
\ee
where indices $a,b$ run over the set $\{0,1,2,\tilde 2,11,21\}$
and the corresponding dimensions are:
\be
d_0=1, \ \ \ d_1 = \frac{D_1D_{-1}}{\{q\}^2}, \ \ \
d_{2} = \frac{D_3D_0^2D_{-1}}{[2]^2\,\{q\}^4}, \ \ \
\tilde d_2 = \frac{D_2D_1D_{-1}D_{-2}}{[2]^2\,\{q\}^4}, \ \ \
d_{11} = \frac{D_1D_0^2D_{-3}}{[2]^2\,\{q\}^4}, \ \ \
d_{21} = \frac{D_3D_1^2D_{-1}^2D_{-3}}{[3]^2\,\{q\}^6}
\nn
\ee
\be
d_{3} = \frac{D_5D_1^2D_0^2D_{-1}}{[3]^2[2]^2\,\{q\}^6}, \ \ \
\tilde d_{3} = \frac{D_4D_2D_0^2D_{-1}D_{-2}}{[3]^2[2]\,\{q\}^6}, \ \ \
d_{31} = \frac{D_5D_2^2D_0^2D_{-1}^2D_{-3}}{[4]^2[2]^2\{q\}^8}
\ee
(dimensions in the second line will matter in sec.6 below).
Multiplicities matter when $a$ or $b$  equals $1$.
Also the "extra" diagrams with the eigenvalue $\tilde\Lambda_2$ appear twice.
In the standard notation  the labeling is different:
\be
\begin{array}{c|cccccc}
{\rm present\ paper} & \ 0 & 1 & 2p & 2 & 2m & 21 \\
\hline
{\rm previous} \  [10,12,32]
& \ 1 & 7-10 & 6 & 2 \,\&\, 3 & 4 & 5
\end{array}
\ee
Accordingly we have the following expressions for the matrix elements
$\bar{\cal S}_{ab}=\bar{\cal S}_{ba}$
through those of the $10\times 10$ symmetric unitary Racah matrix $\bar S_{ij}$,
which was first calculated in \cite{GJ} and then re-deduced by the two different
evolution-based methods in \cite{mmms21} and \cite{rectwist}:
\be
\!\!\!\!\!\!\!\!\!
{\footnotesize
\begin{array}{cccccc}
\bar{\cal S}_{0,0} = \bar S_{11} &
\bar{\cal S}_{0,1} = \sum_{j=7}^{10}\bar S_{1j} &
\bar{\cal S}_{0,2} = \bar S_{16} &
\bar{\cal S}_{0,\tilde 2} = \bar S_{12}+\bar S_{13} &
\bar{\cal S}_{0,11} = \bar S_{14}  &
\bar{\cal S}_{0,21} = \bar S_{15} \\ \\
\bar{\cal S}_{1,0} = \sum_{i=7}^{10} \bar S_{i1} &
\bar{\cal S}_{1,1} = \sum_{i,j=7}^{10}\bar S_{ij} &
\bar{\cal S}_{1,2} = \sum_{i=7}^{10}\bar S_{i6} &
\bar{\cal S}_{1,\tilde 2} = \sum_{i=7}^{10}\bar S_{i2}+\bar S_{i3} &
\bar{\cal S}_{1,11} = \sum_{i=7}^{10}\bar S_{i4}  &
\bar{\cal S}_{1,21} = \sum_{i=7}^{10}\bar S_{i5} \\ \\
\bar{\cal S}_{2,0} = \bar S_{61} &
\bar{\cal S}_{2,1} = \sum_{j=7}^{10}\bar S_{6j} &
\bar{\cal S}_{2,2} = \bar S_{66} &
\bar{\cal S}_{2,\tilde 2} = \bar S_{62}+\bar S_{63} &
\bar{\cal S}_{2,11} = \bar S_{64}  &
\bar{\cal S}_{2,21} = \bar S_{65} \\ \\
\bar{\cal S}_{\tilde 2,0} = \bar S_{21}+\bar S_{23} &
\bar{\cal S}_{\tilde 2,1} = \sum_{j=7}^{10}(\bar S_{2j}+\bar S_{3j}) &
\bar{\cal S}_{\tilde 2,2} = \bar S_{66} &
\bar{\cal S}_{\tilde 2,\tilde 2}
= \bar S_{22}+\bar S_{23}+\bar S_{32} + \bar S_{33} &
\bar{\cal S}_{\tilde 2,11} = \bar S_{24}+\bar S_{34}  &
\bar{\cal S}_{\tilde 2,21} = \bar S_{25} +\bar S_{35} \\ \\
\bar{\cal S}_{11,0} = \bar S_{41} &
\bar{\cal S}_{11,1} = \sum_{j=7}^{10}\bar S_{4j} &
\bar{\cal S}_{11,2} = \bar S_{46} &
\bar{\cal S}_{11,\tilde 2} = \bar S_{42}+\bar S_{43} &
\bar{\cal S}_{11,11} = \bar S_{44}  &
\bar{\cal S}_{11,21} = \bar S_{45} \\ \\
\bar{\cal S}_{21,0} = \bar S_{51} &
\bar{\cal S}_{21,1} = \sum_{j=7}^{10}\bar S_{5j} &
\bar{\cal S}_{21,2} = \bar S_{56} &
\bar{\cal S}_{21,\tilde 2} = \bar S_{52}+\bar S_{53} &
\bar{\cal S}_{21,11} = \bar S_{54}  &
\bar{\cal S}_{21,3} = \bar S_{55} \\ \\
\end{array}
}
\nn
\ee
Note that the $6\times 6$ matrix $\bar{\cal S}$ in (\ref{21calS})
is symmetric, but {\it not} unitary.
The lacking elements of the $10\times 10$ unitary $\bar S$ can be restored
by solving the unitarity constraints,
what once again reproduces the result of \cite{GJ}.

\bigskip

{\bf 5.3.} For {\bf other   defect-zero \cite{Konodef} knots} ${\cal K}^{(0)}$
(when Alexander polynomial in the
fundamental representation is of degree one in $q^{\pm 2}$, i.e.
contains only three-terms)
we expect the differential expansion with the same  $Z$-factors
(\ref{Zf21}), which depend on representation, but not on the knot,
and with different, knot-dependent $F$-factors, i.e.
the expectation is that
\be
{\cal H}_{[2,1]}^{{\cal K}^{(0)}} = 1
+ Z^{[1]}_{[2,1]}\cdot F_{[1]}^{{\cal K}^{(0)}}
+ \frac{[3]}{[2]}\cdot Z^{[2]}_{[2,1]}\cdot
F_{[2]}^{{\cal K}^{(0)}}
+ \frac{[3]}{[2]}\cdot Z^{[1,1]}_{[2,1]}\cdot
F_{[1,1]}^{{\cal K}^{(0)}}
+ Z^{[2,1]}_{[2,1]}\cdot
F_{[2,1]}^{{\cal K}^{(0)}}
+ {\tZ}^{[2,1]}_{[2,1]}\cdot
\tF_{[2,1]}^{{\cal K}^{(0)}}
\label{H21K0}
\ee
with polynomial $F_{[\lambda]}^{{\cal K}^{(0)}}$.
Moreover,  $F_{[1]}^{{\cal K}^{(0)}}$ and
$F_{[1,1]}^{{\cal K}^{(0)}}(A,q)=F_{[2]}^{{\cal K}^{(0)}}(A,-q^{-1})$
are defined from the expansions of simpler colored HOMFLY-PT polynomials
\be
{\cal H}_{[1]}^{{\cal K}^{(0)}}=1+D_1D_{-1}\cdot F_{[1]}^{{\cal K}^{(0)}}\nn \\
 {\cal H}_{[2]}^{{\cal K}^{(0)}}=1+[2]D_2D_{-1}\cdot F_{[1]}^{{\cal K}^{(0)}}
 + D_3D_2D_0D_{-1}\cdot F_{[2]}^{{\cal K}^{(0)}}
\ee
Indeed, for the simplest defect-zero knot, which is not a double braid,
${\cal K}=9_{46}$, we get:
\be
F_{[1]}^{9_{46}} = A^2\cdot (A^2+1) \nn \\
F_{[2]}^{9_{46}} = A^4\cdot (q^8A^4 +[2]q^5A^2 +1) \nn \\
F_{[3]}^{9_{46}} = q^4A^6\cdot(q^4A^2+1)\cdot (q^{16}A^4 +[2]q^9A^2 -q^6+q^2+1)\nn \\
F_{[2,1]}^{9_{46}} = A^6\cdot\left(A^6+\frac{[6]}{[2]}\cdot A^4 + [3]\cdot\Big(
1+[3]\,\{q\}^2+\frac{[6]}{[2]}\,\{q\}^4\Big)\cdot A^2 + (1+[3]\,\{q\}^2)\right) \nn \\
\tF_{[2,1]}^{9_{46}} = A^6\cdot \left(\{q\}^2\cdot A^6-\frac{[8]}{[2]}\,\{q\}^2\cdot A^4
+\frac{[6]}{[2]}([3]+[5]\,\{q\}^2)\cdot A^2 - \frac{[18]}{[9][2]}\right)
\ee
$F_{[3]}$ does not contribute to the expansion (\ref{H21K0}) of ${\cal H}_{[2,1]}$
in representation $[2,1]$,
it is provided here for comparison and for future use.

\bigskip

{\bf 5.4.}
For knots with non-zero defect a slightly weaker form of the DE can be expected,
with underlined differentials omitted from the $Z$-factors in (\ref{Zf21}).
For example, for the knot $6_2$ with defect one
\be
{\cal H}_{[2,1]}^{6_2} = 1+ Z^{[1]}_{[2,1]}\cdot F_{[1]}^{6_2} +
\frac{[3]}{[2]}\,D_2D_{-2}\,
\Big(D_3\cdot G_{[2]}^{6_2} + D_{-3}\cdot G_{[1,1]}^{6_2}\Big)
+D_3D_2D_{-2}D_{-3}\cdot G_{[2,1]}^{6_2}
- [3]^2\,\{q\}^4\,D_2D_{-2}\cdot \tilde G_{[2,1]}^{6_2}
\nn
\ee
with
\vspace{-0.7cm}
\be
G_{[1]}^{6_2} = \frac{[6]}{[3][2]}\,A^2 \nn \\
G_{[2]}^{6_2} = q^2A^4\,\left(D_2 + \frac{[6][2]}{[3]}\,\{q\}^2\,D_0\right)
-[2]\,qA^3\{q\}^2 \nn \\
G_{[2,1]}^{6_2} = A^6D_0^2 + \{q\}^2\,A^4\,\left(\frac{[6]^2}{[2]^2}\,A^4
- \frac{[4]}{[2]}\big([4][2]+[3]\big)\,A^2 + \frac{[6][5]}{[2]}\right) \nn \\
\tilde G_{[2,1]}^{6_2} = A^4\,\left([3]A^4
- [3]\Big(3+2\,\frac{[4][3]}{[2]}\,\{q\}^2\Big)A^2  + \frac{[6]}{[2]}\right)
\label{G62}
\ee

\vspace{-0.7cm}
\section{Representation $R=[3,1]$}

This time the Young diagrams in the product $R\otimes \bar R = [3,1]\otimes
\overline{[3,1]}$  are:

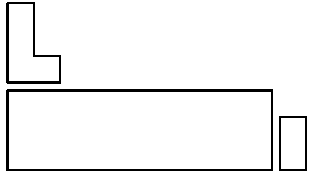
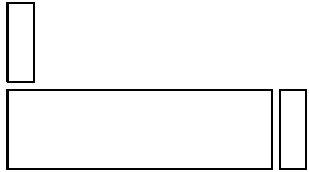
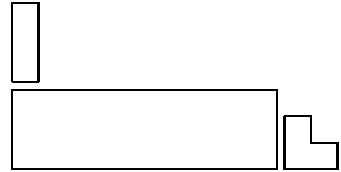
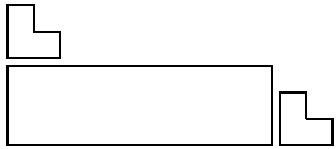
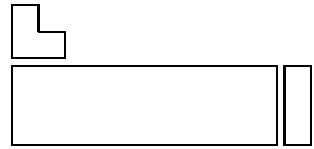
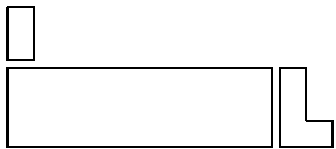
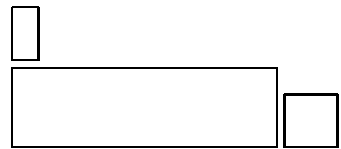
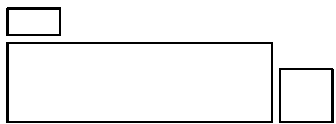
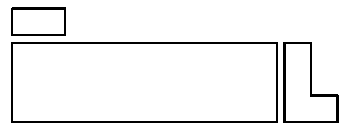
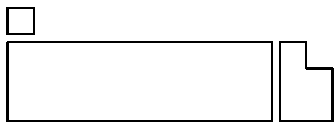
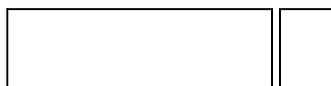
\vspace{-0.2cm}
\begin{picture}(500,650)(-50,-595)
\put(0,-20){\picBdd
\put(0,33){\picde}
\put(103,0){\picz}
\put(30,-20){\put(0,0){\mbox{$\lambda=[3,1]$}}}
}
\put(0,-120){\picBdd
\put(0,33){\picd }
\put(103,0){\picd }
\put(30,-20){\put(0,0){\mbox{$\lambda=[3]$}}}
}
\put(200,-120){\picBdd
\put(0,33){\picd }
\put(103,0){\picze }
\put(30,-20){\put(0,0){\mbox{$\lambda\notin R$}}}
}
\put(0,-220){\picBdd
\put(0,33){\picze }
\put(103,0){\picze }
\put(30,-20){\put(0,0){\mbox{$\lambda=[2,1]$}}}
}
\put(200,-220){\picBdd
\put(0,33){\picze }
\put(103,0){\picd }
\put(30,-20){\put(0,0){\mbox{$\lambda\notin R$}}}
}
\put(0,-320){\picBdd
\put(0,33){\picz }
\put(103,0){\picde }
\put(30,-20){\put(0,0){\mbox{$\lambda=[2]$}}}
\put(-30,10){\mbox{$M$}}
}
\put(200,-320){\picBdd
\put(0,33){\picz }
\put(103,0){\piczz }
\put(30,-20){\put(0,0){\mbox{$\lambda\notin R$}}}
}
\put(0,-410){\picBdd
\put(0,33){\picee }
\put(103,0){\piczz }
\put(30,-20){\put(0,0){\mbox{$\lambda=[1,1]$}}}
}
\put(200,-410){\picBdd
\put(0,33){\picee }
\put(103,0){\picde }
\put(30,-20){\put(0,0){\mbox{$\lambda\notin R$}}}
}
\put(0,-490){\picBdd
\put(0,33){\pice }
\put(103,0){\picdz}
\put(30,-20){\put(0,0){\mbox{$\lambda=[1]$}}}
\put(-30,10){\mbox{$M$}}
}
\put(0,-560){\picBdd
\put(0,33){  }
\put(103,0){\picdd }
\put(30,-20){\put(0,0){\mbox{$\lambda=\emptyset$}}}
}
\end{picture}
The last two diagrams in the right column are exactly the same
as in the case of $R=[2,1]$ (with one full line added at the bottom,
which does not affect dimensions and Casimirs) and therefore should provide the
same single contribution $\tF_{[2,1]}$.
The first two diagrams in the right column
again have coincident dimensions and Casimir eigenvalues
and provide another unified contribution $\tF_{[3,1]}$.

\be
{\cal H}_{[3,1]}^{(m,n)} = 1
+ Z^{[1]}_{[3,1]}\cdot \frac{F_{[1]}^{(m)}F_{[1]}^{(n)}}{F_{[1]}^{(1)}F_{[1]}^{(-1)}}
+   Z^{[2]}_{[3,1]}\cdot
\frac{F_{[2]}^{(m)}F_{[2]}^{(n)}}{F_{[2]}^{(2)}F_{[1]}^{(-1)}}
+ \frac{[4]}{[2]}\cdot Z^{[1,1]}_{[3,1]}\cdot
\frac{F_{[1,1]}^{(m)}F_{[1,1]}^{(n)}}{F_{[1,1]}^{(1)}F_{[1,1]}^{(-1)}}
+\nn \\
+ \frac{[4]}{[3]}\cdot Z^{[3]}_{[3,1]}\cdot
\frac{F_{[3]}^{(m)}F_{[3]}^{(n)}}{F_{[3]}^{(1)}F_{[3]}^{(-1)}}
+ \frac{[4][2]}{[3]}\cdot Z^{[2,1]}_{[3,1]}\cdot
\frac{F_{[2,1]}^{(m)}F_{[2,1]}^{(n)}}{F_{[2,1]}^{(1)}F_{[2,1]}^{(-1)}}
+ Z^{[3,1]}_{[3,1]}\cdot
\frac{F_{[3,1]}^{(m)}F_{[3,1]}^{(n)}}{F_{[3,1]}^{(1)}F_{[3,1]}^{(-1)}} + \nn \\
  + {\tZ}^{[2,1]}_{[3,1]}\cdot
\frac{\tF_{[2,1]}^{(m)}\tF_{[2,1]}^{(n)}}{\tF_{[2,1]}^{(1)}\tF_{[2,1]}^{(-1)}}
+ {\tZ}^{[3,1]}_{[3,1]}\cdot
\frac{\tF_{[3,1]}^{(m)}\tF_{[3,1]}^{(n)}}{\tF_{[3,1]}^{(1)}\tF_{[3,1]}^{(-1)}}
\ee
with the $Z$-factors
\be
Z^{[1]}_{[3,1]}=D_5D_{-3} + D_3D_1+ D_2D_{-2}+D_0D_{-2}, \ \ \ \ \
Z^{[2]}_{[3,1]}= D_3\underline{D_0}\cdot
\Big( D_5D_{-2}+\frac{[6]}{[3]}D_4D_{-2} +D_2D_{-1}\Big), \nn\\
Z^{[1,1]}_{[3,1]}= D_3\underline{D_0}D_{-2}D_{-3}, \ \ \ \
Z^{[3]}_{[3,1]}= D_5D_4D_3\underline{D_1 D_0}D_{-2}, \ \ \ \
Z^{[2,1]}_{[3,1]}= D_4D_3 \underline{D_1 D_{-1}}D_{-2}D_{-3},\nn \\
Z^{[3,1]}_{[3,1]}= D_5D_4D_3D_2D_0D_{-1}D_{-2}D_{-3}, \ \ \ \
\tZ^{[2,1]}_{[3,1]}= -[4]^2[2]\{q\}^4 D_3 D_{-2}, \ \ \ \
\tZ^{[3,1]}_{[3,1]}=  -[4]^2[2]\{q\}^4 D_4D_3D_0D_{-2}
\ee
They are deduced/guessed from the belief in the DE structure
and factorization conjecture, used is also the knowledge \cite{mmms21} of the
$[3,1]$-HOMFLY of the three 3-strand twist knots $3_1,4_1$ and $5_2$.
Underlined are the differentials, which should be omitted in the case of
knots with non-vanishing defects.
Important insight from (\ref{G62}) is that $D_1D_{-1}$ should be obligatory
present in $Z^{[2,1]}_{[3,1]}$ -- to be eliminated in the DE for knots with
non-vanishing defects.
Additional $F$-functions, which did not appear in the list (\ref{Ftwist21}), are:
\be
F_{[3]}^{(m)} = q^3A^3\cdot\left(\frac{\Lambda_0^{2m}}{D_2D_1D_0}
- [3]\cdot\frac{\Lambda_1^{2m}}{D_3D_2D_0}
+ [3]\cdot\frac{\Lambda_{2}^{2m}}{D_4D_2D_1}
- \frac{\Lambda_{3}^{2m}}{D_4D_3D_2}\right)
\nn \\
F_{[3,1]}^{(m)} =  q^2A^4\cdot\left(\frac{\Lambda_0^{2m}}{D_2D_1D_0D_{-1}}
- [4]\cdot \frac{\Lambda_1^{2m}}{D_3D_2D_0D_{-2}}
+ [4]\cdot \frac{\Lambda_{2}^{2m}}{D_4D_2D_1D_{-2}}
+ \frac{[4]}{[2]}\cdot \frac{\Lambda_{11}^{2m}}{D_3D_2D_{-1}D_{-2}}
- \right. \nn \\ \left.
- \frac{[4]}{[3]}\cdot\frac{\Lambda_{3}^{2m}}{D_4D_3D_2D_{-2}}
- \frac{[4][2]}{[3]}\frac{\Lambda_{21}^{2m}}{D_4D_2D_0D_{-2}}
+ \frac{\Lambda_{31}^{2m}}{D_4D_3D_1D_{-2}}\right)
\ee
and
\be
\tF_{[3,1]}^{(m)} = q^2A^4\cdot\left\{
\frac{\Lambda_0^{2m}}{D_2D_1D_0D_{-1}}
 -     \frac{\Lambda_1^{2m}\cdot\big(D_3+[2]D_{-2}\big)}{D_3D_2D_0^2D_{-2}}
\ - \right.\nn \\ \left.
-\frac{1}{[2]^2\{q\}^2}\cdot\left(\frac{\Lambda_{2}^{2m}}{D_4D_2D_1D_{-2}}
\cdot\Big(D_2D_1-\{q\}^2\left([6]+2\cdot[4]\right)\Big)
-  \frac{\tilde\Lambda_2^{2m}\cdot 2D_1}{D_3D_0^2}
+   \frac{\Lambda_{11}^{2m}\cdot D_{-3}}{D_2D_{-1}D_{-2}}
+ \right)\right. \nn \\ \left.
+ \frac{1}{[3]^2\{q\}^2}\cdot\left(\frac{ \Lambda_{3}^{2m}\cdot D_5}{D_4D_3D_2 }
 - 2\frac{\tilde\Lambda_{3}^{2m}}{D_3D_1}
+ \frac{\Lambda_{21}^{2m}\cdot D_{-1}D_{-3}}{D_2D_0^2D_{-2}}\right)
\right\}
\ee
The first two are contributing in the case of rectangular
representations and are known from \cite{rectwist},
the last one is peculiar for non-rectangular case and is new.
Note that, like $\tF_{[2,1]}$ was independent of $\Lambda_{21}$,
this $\tF_{[3,1]}$ does not depend on $\Lambda_{31}$.

The relevant eigenvalues are
\be
\Lambda_0^2 = 1, \ \ \ \Lambda_1^2 = A^2, \ \ \
\Lambda_{2}^2 = q^4A^4, \ \ \ \underline{\tilde\Lambda_2^2 =  A^4}, \ \ \
\Lambda_{11}^2 = q^{-4}A^4, \nn \\
\Lambda_{21}^2 = A^6,\ \ \ \underline{\tilde\Lambda_{3}^2 = q^{6}A^{6}}, \ \ \
\Lambda_{3}^2 = q^{12}A^{6},\ \ \ \Lambda_{31}^2 = q^8A^8
\ee
Underlined are the two eigenvalues, associated with the two pairs of "anomalous"
Young diagrams.
We remind that for each integer $m$ all $F_\lambda^{(m)}$ are polynomials,
moreover, they drastically simplify to  (\ref{Fdist}) for $m=0,\pm 1$.
Factorization conjecture for double braids is checked by the reduction
property (\ref{Alred}) of Alexander polynomials, which is applicable to
the case of $R=[3,1]$.

As in the case of $R=[2,1]$, one can easily deduce the $9\times 9$
matrix $\bar{\cal S}_{ab}$, but it is not unitary, because its elements are actually
averaged over the multiplicity spaces. The elements of the first line are made
from dimensions,
\vspace{-0.3cm}
\be
\bar{\cal S}_{0 b} = \eta_b\frac{\sqrt{d_\emptyset d_b}}{d_{[3,1]}}
\label{S0b}
\ee
with $\eta_b=2$ rather than $1$ for $b=1,2,\tilde 2,\tilde 3$ -- this fact is
actually used in the derivation of $\tF_{[3,1]}$.
These $\eta_b$ (actually, $\eta_b^{[3,1]}$)
take into account the pairwise "degeneracy" of the "anomalous diagrams
$\tilde 2$ and $\tilde 3$ and the multiplicity in the channels $1$ and $2$.
Note that in the case of $R=[2,1]$
the two factors $\eta_1^{[2,1]}=\eta_{\tilde 2}^{[2,1]}=2$ were also non-trivial,
but $\eta_{2}^{[2,1]}=1\neq \eta_{2}^{[3,1]}=2$.
Even in the simplest case of the first line instead of unitarity we have
\vspace{-0.3cm}
\be
\sum_{b=0}^8 \frac{1}{\eta_b}\bar{\cal S}_{0b}^2 = 1
\ \ \Longleftrightarrow \ \  \sum_{b=0}^8 \ \eta_b\cdot d_b = d_{[3,1]}^2
\label{0line}
\ee
with non-trivial $\eta$-weights.
Extraction of the elements of the bigger unitary
exclusive Racah matrix $\bar S$ requires additional effort.
Due to peculiar properties (many vanishing entries) of $\bar S$,
in addition to (\ref{0line}) there are two more elementary sum rules,
involving not only the first, but also the last lines in $\bar{\cal S}$:
$\sum_{b=0}^5 \bar{\cal S}_{21,b}^2 = 1$ and
$\sum_{b=0}^5 \frac{1}{\eta_b}\,\bar{\cal S}_{0,b}\,\bar{\cal S}_{21,b} = 0$
in the case of $R=[2,1]$ and
$\sum_{b=0}^8 \bar{\cal S}_{31,b}^2 = 1$ and
$\sum_{b=0}^8 \frac{1}{\eta_b}\,\bar{\cal S}_{0,b}\,\bar{\cal S}_{31,b} = 0$
in the case of $R=[3,1]$ (note that the first sums do not contain
$\eta_b$-factors).
Like elements in the first line, those in the last are also fully factorized.
Since $\bar{\cal S}$ is symmetric, the same is true about the first and the
last rows.
These boundary elements can be directly identified (modulo factors 2 and 0)
with the  corresponding elements of the unitary $\bar S$.
Like in the case of $R=[2,1]$, all other elements can be restored
by solving the unitarity constraints.
Then they can be compared with the inclusive ones, found for the
case of $R=[3,1]$ in the second paper of \cite{mmms21}.

\section{Other representation $R=[r,1]$}

It is now straightforward to describe the shape of the DE for other non-rectangular
representations, at least for the entire family $R=[r,1]$:
\be
{\cal H}^{(m,n)}_{[r,1]} = 1 +
\sum_{i=1}^r \left( \frac{1}{[r]}\cdot \frac{[r+1]!}{[i]![r-i]!}\cdot
\frac{D_{r+i-2}!\underline{D_{i-2}!}}{D_{r-1}!}
\cdot\Big(D_{r-2}D_{i-1}-[r+1][i]\cdot\{q\}^2\Big)
\cdot\frac{F_{[k]}^{(m)}F_{[k]}^{(n)}}{F_{[k]}^{(1)}F_{[k]}^{(-1)}}
+ \right. \nn \\ \left.
+ \frac{[i]}{[r]}\cdot \frac{[r+1]!}{[i+1]![r-i]!}\cdot
\frac{D_{r+i-1}!D_{i-1}!}{D_{r-1}!\cdot D_{i-2}}
\cdot D_{-1}D_{-2}D_{-3}
\cdot \frac{F_{[k,1]}^{(m)}F_{[k,1]}^{(n)}}{F_{[k,1]}^{(1)}F_{[k,1]}^{(-1)}}\right)
- \ \ \ \ \ \ \ \nn \\
 - \ \{q\}^4\cdot \sum_{i=2}^r   \frac{[r+1]}{[r]}\cdot\frac{[r+1]!}{[r-i]![i-2]!}
\cdot \frac{D_{r+i-2}!D_{i-3}!}{D_{r-1}!}\cdot D_{-2}
\cdot \frac{\tF_{[k,1]}^{(m)}\tF_{[k,1]}^{(n)}}{\tF_{[k,1]}^{(1)}\tF_{[k,1]}^{(-1)}}
\ \ \ \ \ \ \ \ \ \ \ \ \ \ \ \ \ \ \ \ \ \ \ \ \ \ \
\label{Hr1}
\ee
where
$F$-functions are defined in eq.(37) of \cite{rectwist}
(see also \cite{konotwist} for a more profound description
in terms of shifted skew-characters):
\be
F_{[a+1,1^{b}]}^{(m)} = \Big( q^{\frac{a-b}{2}}A\Big)^{a+b+1}\left\{
\frac{1}{\{Aq^a\}\ldots \{A/q^b\}}  \
+ \ \ \ \ \ \ \ \ \ \ \ \ \ \ \ \ \ \ \ \ \ \ \ \ \ \ \ \ \ \ \ \ \ \
\phantom{5^{\int^{\int^{^{\int^{\int^{ \int^5}}}}}}}
\ \ \ \ \ \ \ \ \ \ \ \ \ \ \ \ \ \ \ \   \right. \nn \\ \left.
+ \ \sum_{i=0}^a\sum_{j=0}^b \frac{(-)^{i+j+1}\Big(q^{i-j}A\Big)^{2m\cdot(i+j+1)}
\frac{[a]!}{[a-i]![i]!}\cdot\frac{[b]!}{[b-j]![j]!}\cdot \frac{[a+b+1]}{[i+j+1]}}
{\{Aq^{a+i+1}\} \ldots\{Aq^{i+1}\} \cdot \frac{\{Aq^{i-j}\}}{\{Aq^{2i+1}\}\{A/q^{2j+1}\}}
\cdot \{A/q^{j+1}\}\ldots \{A/q^{b+j+1}\}}
\right\}
\ee
and $\tF$-functions can be restored recursively in $r$ from the conditions
(\ref{S0b}),  as explained in the previous section,
 -- see eq.(\ref{tF}) below.
As usual, they vanish for the unknot ($m=0$), turn to unity for the figure-eight knot
$4_1$ ($m=-1$) and coincide with the ordinary $F_{[k,1]}$ for the trefoil $3_1$ ($m=1$):
\be
\tF_{[k,1]}^{(-1)}=F_{[k,1]}^{(-1)} = 1  \ \ \ \ \ \ \ \ \ \ \ \ \nn \\
\tF_{[k,1]}^{(0)}=F_{[k,1]}^{(0)} = 0 \ \ \ \ \ \ \ \ \ \ \ \ \ \ \nn \\
\tF_{[k,1]}^{(1)}=F_{[k,1]}^{(1)} = (-)^{k+1}\cdot A^{2k+2}\cdot q^{(k-2)(k+1)}
\label{Fdistin}
\ee
However, for other $m$ there is no coincidence: $\tF_{[k,1]}^{(m)}\neq F_{[k,1]}^{(m)}$.
Note that $D$-factorials $D_n! = \prod_{i=0}^n D_i$ are defined as products
of $n+1$ differentials, beginning from $i=0$.

Dimensions, needed to determine $\tF$-functions are equal to
\be
d_i =  \frac{D_{2i-1}\,\big(D_{i-2}!\big)^2\,D_{-1}}{\big([i]!\big)^2\{q\}^{2i}},\ \ \ \
d_{i1} = \frac{D_{2i-1}\,\big( D_{i-1} D_{i-3}! D_{-1}\big)^2\,D_{-3}}
{\big([i+1]\,[i-1]!\big)^2\cdot\{q\}^{2i+2}}, \ \ \ \
\tilde d_i = \frac{D_{2i-2}D_{i-1}\,\big(D_{i-3}!\big)^2\,D_{-1}D_{-2}}
{[i][i]![i-2]!\cdot\{q\}^{2i}}
\ee
and satisfy
\be
d_0 + 2\cdot \sum_{i=1}^{r-1} d_i + d_r + \sum_{i=1}^r d_{i1} + 2\cdot \sum_{i=2}^r
\tilde d_i \ = \ d_{[r,1]}^2 \ =
\ \left(\frac{D_{r-1}!\,D_{-1}}{[r+1]\,[r-1]!\cdot \{q\}^{r+1}}\right)^2
\ee
what defines the coefficients $\eta_b$.
Non-trivial multiplicities $\eta_i$ arise for $i=1,\ldots,r-1$,
while $\tilde \eta_{ i}=2$ for $\tilde i = 2,\ldots,r$
account just for the two copies of  "anomalous" diagrams in
decomposition of $[r,1]\otimes\overline{[r,1]}$, which do not mix.
Associated eigenvalues are
\be
\Lambda_i^2=q^{2i(i-1)}A^{2i},  \ \ \ \ \ \
\Lambda_{i1}^2 = q^{2(i+1)(i-2)}A^{2i+2}, \ \ \ \ \ \
\tilde\Lambda_i^2 = q^{2i(i-2)} A^{2i}
\ee
In these terms
\be
\tF_{[k,1]}^{(m)} = q^{\frac{(k+1)(k-2)}{2}}A^{k+1}\cdot\left\{
\frac{\Lambda_0^{2m}}{\prod_{l=-1}^{k-1}D_l}
 -     \frac{\Lambda_1^{2m}\cdot\big(D_k+[k-1]D_{-2}\big)}
 {  D_0^2D_{-2}\cdot \prod_{l=2}^k D_l}
\ +  \right.
\label{tF}
\ee
\vspace{-0.5cm}
\be
 \left.\!\!\!\!\!\!\!\!\!\!\!\!\!\!\!\!\!
+ \sum_{i=2}^{k} \frac{(-)^{i+1}}{[i]^2\,\{q\}^2 }
\cdot\frac{[k-2]!}{[i-2]![k-i]!}
\cdot\frac{ \Lambda_{i}^{2m}
\cdot D_{2i-1}\cdot D_{i-2}!  }
{  D_{i-1} D_{-2}\cdot  D_{k+i-1}!}
\cdot  \left(D_{k+i-1}D_{k-1}D_{-2}
-\{q\}^2\cdot\frac{[k-i]}{[i-1]}\Big(D_{k+i-1}+[k-1][i]\cdot D_{-2}\Big)
\right)
+ \right.\nn \\  \left. \!\!\!\!\!\!\!\!\!\!\!\!\!\!\!\!\!
+ \sum_{i=1}^{k-1} \frac{(-)^i}{[i+1]^2\,\{q\}^2 }
\cdot\frac{[k-2]!}{[i-1]![k-i-1]!}\cdot\frac{ \Lambda_{i1}^{2m}
\cdot D_{2i-1} D_{-1}D_{-3}}
{  D_{i-2}^2D_{-2}\cdot  \prod_{l=i}^{k+i-2} D_l}
+ 2\cdot \sum_{i=2}^{k} \frac{(-)^i}{[i]^2\,\{q\}^2 }
\cdot\frac{[k-2]!}{[i-2]![k-i]!}\cdot\frac{ \tilde\Lambda_{i}^{2m}
\cdot D_{k-2}D_{2i-2}  }
{  D_{i-2}^2D_{i}\cdot \prod_{l=i+1}^{k+i-2} D_l }
 \right\}
\nn
\ee

 \noindent
 Eq.(\ref{Hr1}) successfully reproduces the only available answers
in all representations inside the double-braid family --
those for the trefoil $3_1$.

\section{Towards exclusive Racah matrix $\bar S$ for $R=[r,1]$}

As already mentioned, the suggestion of \cite{rectwist}
was  to extract exclusive Racah matrices from the $\Lambda$-expansion of
the antiparallel double braids, see (\ref{21calS}) in sec.5 above:
\be
H_{R}^{(m,n)} = \sum_{a,b}
\frac{\sqrt{d_ad_b}}{d_{R}}\cdot \bar {\cal S}_{ab}
\cdot \Lambda_a^{2m}\Lambda_b^{2n}
\label{calS}
\ee
This is an example of the evolution-based \cite{evo} approach to Racah calculus,
proposed and successfully used in \cite{mmms21}.
Factorization of differential expansion for double braids, which
allows to efficiently calculate $H_{R}^{(m,n)}$,
opens spectacular possibilities for the case of exclusive Racah $\bar S$.
The only remaining problem is that for non-rectangular representations
eq.(\ref{calS}) contains the "averaged"  matrix $\bar{\cal S}$,
which is non-unitary and smaller than the unitary matrix $\bar S$ itself.
We now describe a way to restore $\bar S$ from the known $\bar{\cal S}$.
Basically, it uses unitarity constraints to find the lacking elements
of the bigger matrix.
With known $\bar{\cal S}$, the number of conditions for $\bar S$
seems sufficient.
However, one could expect that this approach is impractical,
because it involves solving
a number of quadratic relations, involving polynomials of
complexity, which fastly increases with $r$.
Fortunately, things turn to be a little simpler.
In the by-now-standard case of $R=[2,1]$
the testable properties of $\bar{\cal S}$
allow a procedure, which  involves solving just {\it linear} equations and
taking a few square roots of the factorized expressions.
For $r>2$ the situation is more involved, still
calculation looks {\it practically possible}.

\bigskip

The matrix $\bar{\cal S}$, which can be directly read from (\ref{calS}),
has dimension $3r\times 3r$ and is symmetric,
but not unitary.
The unitary Racah matrix $\bar S$ has the lines/columns $\tilde i$ doubled
and $i$ with $0<i<r$ quadrupled -- see \cite{mmmrsv1} for detailed
explanations and comments.
In result this symmetric and unitary matrix has the size $ (7r-4) \times (7r-4) $.
If we keep the order of columns, chosen in \cite{mmmrsv1} for the case of
$R=[2,1]$
(with second and the third columns permuted to make the matrix symmetric),
then the $ (7r-4) \times (7r-4) $ symmetric and unitary $\bar S$ is expressed
through the $3r\times 3r$ symmetric, but not unitary, $\bar {\cal S}$
as follows:

\be
\!\!\!\!\!\!\!\!\!\!\!\!\!\!\!\!\!\!\!\!\!\!\!\!\!\!
\bar S =\left(
\begin{array}{c||c|cc|ccc|cccc}
& & \tilde j = \!\!\!\!\!\!\!\!\!\!\!\!\!\!\!&\!\!\!\!\!\!\!\!\!\!\!\!\!\!\! 2,\ldots,r
& j=1,\ldots,r-1 & &&& j=\!\!\!&1,\ldots,r-1 &\\
&&&&&&&&&&\\
& 0 & \tilde j,1 & \tilde j,2  & j1 & r1 & r & j,1 & j,2 & j,3 & j,4 \\
&&&&&&&&&&\\
\hline \hline
&&&&&&&&&&\\
0 & \BS_{00} & \frac{1}{2}\BS_{0\tilde j} & \frac{1}{2}\BS_{0\tilde j}
& \BS_{0,j1} & \BS_{0,r1} & \BS_{0r} & \frac{1}{2}\BS_{0j}
& \frac{1}{2}\BS_{0j} & 0 & 0 \\
&&&&&&&&&&\\
\hline
&&&&&&&&&&\\
\tilde i,1 & \frac{1}{2}\BS_{0,\tilde i}
&  \frac{1}{4}\BS_{\tilde i\tilde j} &  \frac{1}{4}\BS_{\tilde i\tilde j}
&  \frac{1}{2}\BS_{\tilde i,j1} & 0 & \frac{1}{2}\BS_{\tilde i,r}
&\frac{1}{2}\BS_{\tilde i\,j}-x_{ij} & x_{ij}
& \frac{Y_{ij}+y_{ij}}{2} & \frac{Y_{ij}-y_{ij}}{2}
\\
&&&&&&&&&&\\
\tilde i,2 & \frac{1}{2}\BS_{0,\tilde i}
&  \frac{1}{4}\BS_{\tilde i\tilde j} &  \frac{1}{4}\BS_{\tilde i\tilde j}
&  \frac{1}{2}\BS_{\tilde i,j1} & 0 & \frac{1}{2}\BS_{\tilde i,r}
&\frac{1}{2}\BS_{\tilde i\,j}-x_{ij} & x_{ij}
& \frac{-Y_{ij}+y_{ij}}{2} & \frac{-Y_{ij}-y_{ij}}{2}
\\
&&&&&&&&&&\\
\hline
&&&&&&&&&&\\
i1 & \BS_{0,i1} & \frac{1}{2}\BS_{i1,\tilde j} & \frac{1}{2}\BS_{i1,\tilde j}
& \BS_{i1,j1} & \BS_{i1,r1} & \BS_{i1,r}
& \BS_{i1,j}-u_{ij} & u_{ij} & v_{ij} & -v_{ij}
\\
&&&&&&&&&&\\
r1 & \BS_{0,r1} & 0 & 0 & \BS_{i1,r1} & \BS_{r1,r1} & \BS_{r1,r} &
\BS_{r1,j} & 0 & 0 & 0
\\
&&&&&&&&&&\\
r & \BS_{0r} & \frac{1}{2}\BS_{r,\tilde j} & \frac{1}{2}\BS_{r,\tilde j}
& \BS_{r,j1} & \BS_{r,r1} & \BS_{r,r}
& \BS_{rj}-U_{j} & U_{j} & V_{j} & -V_{j}
\\
&&&&&&&&&&\\
\hline
&&&&&&&&&&\\
i,1 & \frac{1}{2}\BS_{0i} & \frac{1}{2}\BS_{i,\tilde j}-x_{ji}
& \frac{1}{2}\BS_{i,\tilde j}-x_{ji}
& \BS_{i,j1}-u_{ji} & \BS_{i,r1} & \BS_{ir}-U_i
& z_{ij|11} & z_{ij|12} & z_{ij|13} & z_{ij|14}
\\
&&&&&&&&&&\\
i,2 & \frac{1}{2}\BS_{0i} & x_{ji} & x_{ji} & u_{ji} & 0 & U_i
& z_{ij|21} & z_{ij|22} & z_{ij|23} & z_{ij|24}
\\
&&&&&&&&&&\\
i,3 & 0 & \frac{Y_{ij}+y_{ij}}{2} & \frac{-Y_{ij}+y_{ij}}{2} & v_{ji} & 0 & V_i
& z_{ij|31} & z_{ij|32} & z_{ij|33} & z_{ij|34}
\\
&&&&&&&&&&\\
i,4 & 0 & \frac{Y_{ij}-y_{ij}}{2} & \frac{-Y_{ij}-y_{ij}}{2} & -v_{ji} & 0 & -V_i
& z_{ij|41} & z_{ij|42} & z_{ij|43} & z_{ij|44}
\\
&&&&&&&&&&\\
\end{array}
\right)
\nn
\ee

\bigskip

\noindent
Remaining parameters $x,u,v,z$ are not defined from the double-braid
evolution.
However, they are unambiguously (up to inessential signs)
dictated by the unitarity of the matrix $\bar S$.
Calculation are greatly simplified (almost reduced to just {\it linear} equations,
with most quadratic left only for the checks of consistency)
because of the special property of the distinguished line/column
with $r1$, associated with the "biggest" diagram in $R\otimes \bar R$
 -- for this reason it is written separately from all other $i1$.
Numerous zeroes in this case (and also in the very first line)
are implied by the three identities, already familiar from the end of sec.6:
\be
\sum_{b=0}^{3r-1} \frac{1}{\eta_b}\bar{\cal S}_{0b}^2 = 1
\ \ \Longleftrightarrow \ \  \sum_{b=0}^{3r-1} \ \eta_b\cdot d_b = d_{[r,1]}^2
 \ \
\nn \\
\sum_{b=0}^{3r-1} \bar{\cal S}_{r1,b}^2 = 1,
\ \ \ \ \ \ \ \ \ \ \ \ \ \
\sum_{b=0}^{3r-1} \frac{1}{\eta_b}\,\bar{\cal S}_{0,b}\cdot\bar{\cal S}_{r1,b} = 0
\ee
The fact that the entries $\pm v$ in the last two columns for the lines $i1$ and $r$
differ only by a sign follow from additional properties of $\BS$.

Orthogonality to the two distingushed lines $0$ and $r1$, provides linear equations
for parameters $x$ and $u$ and most of $z$, which for $r=2$
fix them unambigously.
After that $v$'s can be defined by taking the square roots
\be
v_{\tiny\bullet} =
\sqrt{\frac{1}{2}\left(1-\sum_{J=1}^{7r-6} {\bar S}_{{\tiny\bullet},J}^2\right)}
\ee
-- this is the only place where the sign ambiguity occurs.
Finally, orthogonality to these lines with known $v$'s provide linear equations
for the remaining $y,Y$ and $z$.
For $r\geq 3$ some bilinear relations  also need to be used.

Once $\bar S$ is known, the second exclusive  $S$
(which is unitary, but not symmetric) can be found
as its diagonalization matrix from
\be
\bar S = \bar T^{-1} ST^{-1} S^\dagger \bar T^{-1}
\ee
with the known diagonal $T$ and $\bar T$,
see \cite{mmmrsv1} and eq.(2) in \cite{rectwist}.

\section{Conclusion}

This paper describes a new important progress in the study of colored knot polynomials.
The form of the differential expansion is finally fixed for the family of
twist knots in the case of the simplest non-rectangular representations
$R=[2,1]$, $R=[3,1]$ and, conjecturally, $R=[r,1]$,
thus complementing the recent results of \cite{rectwist,konotwist} for rectangular $R$.
The crucial feature of the differential expansion is its factorization
for the antiparallel double braid family -- and the main result of
this paper is that it continues to hold for non-rectangular $R$.
As a byproduct we get $[3,1]$-colored HOMFLY-PT polynomials for the
infinite double-parametric double-braid family of defect-zero knots --
of which only three examples
(for the 3-strand ${\cal K}^{(0)}=3_1,4_1,5_2$) were known so far.
Infinitely many new (double-braid, but not twist, known since \cite{twevo21})
are also added in the better-studied $[2,1]$ case.
Conjecture (\ref{Hr1}) for $R=[r,1]$ provides much more new results and has
other far-going implications.

The three immediate next questions to address are
the derivation of general formulas
for the exclusive Racah matrices $\bar S$ and $S$
as functions of $r$,
the search for equations {\it a la} \cite{Gar,knotpolseqs} in $r$
and extension from the   cases of rectangular $R=[r^s]$  and
non-rectangular $R=[r,1]$,
tamed respectively in \cite{rectwist,konotwist} and in the present paper,
to generic representations $R$ .

\section*{Acknowledgements}

This work was performed at the Institute for the
Information Transmission Problems with the support from
the Russian Science Foundation, Grant No.14-50-00150.

\end{document}